# Fourier Domain Physics Informed Neural Network

JONATHAN MUSGRAVE,[1] AND SHUWEI HUANG[1,*]

*Department of Electrical, Computer and Energy Engineering, University of Colorado Boulder, Boulder, Colorado 80309, USA*
*[*]shuwei.huang@colorado.edu*

**Abstract:** Ultrafast optics is driven by a myriad of complex nonlinear dynamics. The ubiquitous presence of governing equations in the form of partial integro-differential equations (PIDE) necessitates the need for advanced computational tools to understand the underlying physical mechanisms. From the experimental perspective, signal-to-noise ratio and availability of measurable data, accounts for a bottle neck in numerical and data-driven modeling methods. In this paper we extend the application of the PINNs architecture to include prior knowledge in both the physical and Fourier domain. We demonstrate our Fourier Domain PINN (FD-PINN) in two distinct forms. The Continuous time FD-PINN predicts accurate solutions to a Generalized Pulse Propagation Equation, which includes the complete delayed nonlinear response, in the data-starved and noisy regime. We extend the architecture to the Discrete time FD-PINN to recover the delayed-response physics from spatially separated measurement points. Our architecture ensures high fidelity predictive modeling and hidden physics recovery for applications such as image reconstruction, pulse characterization and shaping, as well as hidden parameter discovery. The benefits of the FD-PINN for ultrafast nonlinear optics make it immediately experimentally deployable. FD-PINN represents the next generation of tools to study optical phenomena both through modeling and measurements for both forward and inverse problems.

## 1. Introduction

Deep learning is an encompassing term which describes algorithms for automated prediction and decision-making without explicit programmed machine instructions. Today, deep learning has been widely adopted in the field of optics and photonics due to its strength in classification, pattern recognition, prediction, parameter optimization, and the construction of digital models from observable data. Early application of machine learning mainly implemented genetic algorithms for pattern recognition[1], image denoising,[2] aberration correction,[3] reconstruction,[4] pulse characterization, and optical component design.[5,6] Deep learning's superior ability to classify data and identify hidden structures in systems with a large number of degrees of freedom has enabled countless application in photonics and optics.[7]

Recent works have targeted the analysis of larger datasets,[8] solving inverse problems,[9–11] laser design and optimization,[12,13] and countless applications for signal processing and prediction.[7,14] These deep learning networks has seen particular success in machine-aided design of nano-structures and metamaterials,[15,16] control of coherent systems,[17,18] super resolution imaging,[20–22] quantum optics,[22–25] and optical communications.[26]

As the application of deep learning to photonics has become increasingly more sophisticated, a variety of network architectures have proven to be invaluable in predictive and inverse modeling. Feed-forward networks (FFN) excel for applications where input and output connections lack temporal context. Such as in single pass optical systems where FNN has been applied for nonlinear pulse propagation, shaping, and characterization.[27–29] Convolutional neural networks (CNN), a subset of FNN's, excel at reducing data dimensionality such as in multimode fiber dynamics.[30–32] Additionally, temporal/sequential data may be handled by recurrent neural networks (RNN) which have been employed to predict nonlinear pulse dynamics.[33] Deep learning can even be utilized to study basic properties of physical systems. Hidden physics models have emerged where closed-form equations are automatically identified by interpreting samples of dynamic data sets.[34,35] Although this list is not exhaustive, these architectures excel at deriving "black-box" predictive models from large datasets.

However, in ultrafast photonics, the cost of data acquisition is prohibitively expensive for these types of data driven architectures. In these data-starved regimes it is difficult to effectively utilize the previously mentioned architectures in the experimental setting. Particularly because experimental data contains unavoidable noise, training these data-driven deep learning models on a sparse noisy and potentially high dimensional data becomes improbable.

To address these challenges, Physics-Informed Neural Networks[34] (PINNs) combined deep learning with physical models to approximate solutions to governing equations and predict system evolution. PINNs leverage neural networks as universal function approximators,[36] and therefore avoid the need for local time-stepping, linearization, or prior assumptions. Embedding prior knowledge into deep learning architecture makes PINNs particularly robust in data-starved and noisy environments while also enabling these "black-box" neural networks to be interpretable i.e. white box or translucent networks. For these reasons PINN's is an excellent candidate for real-world experimental application where noise is prevalent, and data is sparse. Application of the PINN framework in optics has seen a great deal of success in prediction of nonlinear pulse propagation,[37,44-46] parameter estimation,[38–40] and inverse design.[41,42]

However, its application in few-cycle ultrafast nonlinear optics has not been demonstrated because the PINN's framework is fundamentally incompatible with partial integro-differential equations (PIDEs), a ubiquitous type of governing equation for few-cycle ultrafast nonlinear optics.[43,44] In addition, PIDEs govern a wide breadth of complex phenomena in optics such as multimode nonlinear fiber optics,[45–48] supercontinuum generation,[49] gain managed nonlinear amplification,[50,51] laser-plasma interactions,[52,53] and electromagnetic coupling in nonlinear media.[44] Generally speaking, PIDEs combine the properties of differential equations which account for local changes in the system with integral equations that capture non-local interactions with inhomogeneities. Thus this type equation is not only ubiquitous in optical science but also to biological science,[54,55] epidemiology,[56,57] ecology,[58] economics,[59] and a variety of stochastic processes.[60,61]

In this work we present a novel Fourier domain PINN (FD-PINN) framework specifically designed for deep learning applications for prediction and discovery of systems governed by PIDEs. To enable efficient integration of integro-differential operators into the training process, we specifically target coordinates in the spatiotemporal domain to incorporate governing physics in both the physical and Fourier domain. This extends the PINN's framework to the Fourier domain and incorporates a broad breadth of governing physics. In this paper we focus on studying nonlinear pulse propagation. Previous demonstration of PINNs for nonlinear pulse propagation[38,44-46] required a reduction of the governing PIDE to a pure partial differential equation (PDE) and thus cannot accurately predict the nonlinear pulse propagation in the few-cycle femtosecond regime. Continuous-time FD-PINN (CFD-PINN) solves the problem.

We then extend the FD-PINN architecture to a discrete FD-PINN architecture (DFD-PINN) to tackle hidden continous physics discovery from a system where the physics may not be particularly well known. We harmoniously integrate an implicit Runge-Kutta method (RKM) with the FD-PINN, to discover the Raman response from just one set of input and output (spatially sparse) noisy field data. Previous PINN parameter estimation[38–40] relied on the continuous-time PINN architecture which requires a scattered set of data sampled across the full spatiotemporal domain. Discrete FD-PINN (DFD-PINN) solves the problem.

Finally, we demonstrate that both prediction and discovery algorithms are particularly robust to noise due to the inclusion of the governing PIDE. From the experimental perspective, we provide the first systematic study of how the captured data's signal to noise ratio (SNR) affect the fidelity of the trained models specifically in the low SNR regime. The CFD-PINN and DFD-PINN present the next steps in experimental integration of deep learning for unsupervised approaches to physics discovery and prediction.

The manuscript is arranged as follows; The CFD-PINN is introduced in section II to predict spatiotemporal pulse propagation with sparse labeled data. Section III demonstrates the CFD-PINN's signal recovery in low SNR regimes. Section IV introduces the DFD-PINN which we use to recover the Raman response function from sparse spatial data. Section V demonstrates this inversion algorithm's performance in the low SNR regime. Finally, Section VI concludes.

## 2. Continuous Fourier Domain Physics Informed Neural Network

The generalized pulse-propagation equation, in normalized units, takes the form

$$F_h = h_z - i\frac{D}{2}h_{tt} + \frac{\Delta}{6}h_{ttt} - iN^2(1+is\partial_t)h\int_{-\infty}^{\infty} R(t)|h(t-t')|^2 dt'. \tag{1}$$

Here, $h(t,z)$ is the hidden solution and $F_h$ is the residual of the PIDE and converges to zero once the hidden solution is found. The temporal derivatives, $h_{tt}$ and $h_{ttt}$, represent the second and third order dispersion parametrized by $D$ and $\Delta$ respectively. Together with the spatial derivate these terms comprise the linear operator of the propagation equation. The last term in the equation describes the optical nonlinearities where $N$ is the soliton number and $s$ is the normalized self-steepening parameter. The nonlinear response function $R(t) = (1-f_R)\delta(t) + f_R r(t)$ includes both the instantaneous electronic response defined by the Dirac delta function, $\delta(t)$, The delayed molecular Raman response, $r(t)$, is defined in section 1 of supplemental document. $f_R$ is the fractional contribution of the delayed Raman response to the total nonlinear response. Plugging in the expression of the nonlinear response function, Eq 1 can be rewritten as

$$F_h = h_z - i\frac{D}{2}h_{tt} + \frac{\Delta}{6}h_{ttt} - iN^2(1-f_R)h|h|^2 + N^2(1-f_R)s(h|h|^2)_t \tag{2}$$
$$- iN^2 f_R(1+is\partial_t)h\int_{-\infty}^{\infty} r(t)|h(t-t')|^2 dt'.$$

The fourth term is the self-phase modulation, the fifth term is the self-steeping, and the last term is the convolution integral that describes the intrapulse Raman scattering.

Eq 2 is the PIDE that describes ultrashort pulse propagation down to the few-cycle regime. For longer pulses above 1 ps, one can use the first moment approximation[43] to further simplify it to the generalized nonlinear Schrodinger equation (GNLSE):

$$F_h = h_z - i\frac{D}{2}h_{tt} + \frac{\Delta}{6}h_{ttt} - iN^2(1-f_R)h|h|^2 + N^2(1-f_R)s(h|h|^2)_t \tag{3}$$
$$- \tau_R h(|h|^2)_t.$$

However, as exemplified in Fig. S1, the GNLSE cannot accurately describe the Raman effect in the femtosecond pulse propagation. The shorter the pulse, the larger the discrepancy between the full solution from the PIDE (Eq 2) and the approximated solution from the GNLSE which is a pure PDE (Eq 3). On the other hand, the convolution integral of Eq 2 renders it particularly difficult to solve compared to a PDE where the automatic differentiation in neural network can be directly applied. In the FD-PINN, the PINN's framework is extended to tackle the PIDE by handling the convolution integral in the Fourier domain before it is transformed back to the time domain.

The architecture of the continuous FD-PINN's (CFD-PINN), is illustrated in Fig. 1a. To train the model, the labeled coordinates, spatiotemporal coordinates where the solution the PIDE is known, are fed to the CFD-PINN. At these labeled coordinates, the model predicts the real and imaginary components of the solution defined $h = u + iv$. These predictions are compared to the labeled data. The labeled data are values which can be either experimentally measured or analytically defined and are each individually associated with a labeled coordinate. Loss between predicted field at the labeled coordinate and the labeled data constrain the prediction space of the network to the known values of the hidden solution. We want to emphasize that the prediction of the network does not take any spatial, temporal, or transforms of these predictions thus the labeled coordinates can be seen as a set of independent points meant to constrain the network and does not require knowledge of the system physics or any specific sampling mesh size.

Similarly, the boundary values can be enforced during the training by including boundary coordinates with known boundary data. For localized solutions as those in this paper, the boundary coordinates can be set to zero.

To enforce the physics in the CFD-PINN, collocation coordinates are digitally defined at spatiotemporal points over the domain of interest. These collocation coordinates are specifically selected to sample the domain in a temporal series, defined: $(t_c^i, z_c^j), i = [1, N_z]$ and $j = [1, N_t]$ where $N_z$ and $N_t$ are the total number of spatial and temporal collocations respectively. During the training process, the nonlinear operator is calculated by executing the Fourier transform across the temporal dimension of the prediction made at the collocation coordinates. The Fourier domain prediction is used to calculate the system's response in the frequency domain. The spatial and temporal derivates are found via automatic differentiation and the linear and nonlinear operators are combined to calculate the residual on the PIDE prediction, $F_h$. We want to emphasize that collocation coordinates do not require any labeled data and are only used to train the network on the known physics of the system.

The CFD-PINN we trained in Fig. 1 was defined with a four-layer multioutput deep neural network architecture with 100 neurons to predict the evolution of a 50-fs second order soliton in a standard single mode fiber with the parameters summarized in Section 1 of the supplemental document. The network was trained using 64 labeled coordinates, 80 boundary coordinates, and 20480 collocation coordinates divided into 512 temporal and 40 spatial collocations. Without loss of generality, we assume to have labeled data only at points at the initial condition, $z = 0$.

In this way, the network can efficiently learn the hidden solution by minimizing the root mean squared error of the PIDE, labeled data, and initial/boundary conditions. The loss function of the CFD-PINN is expressed:

$$Loss = MSE_c + MSE_0 + MSE_b \tag{4}$$
$$= \frac{1}{N_z N_t} \sum_{j=1}^{N_z} \sum_{i=1}^{N_t} \left|F_h(t_c^i, z_c^j)\right|^2 + \frac{1}{N_0} \sum_{i=1}^{N_0} \left|h(t_0^i, 0) - h_0^i\right|^2$$
$$+ \frac{1}{N_b} \sum_{i=1}^{N_b} \left|h(t_b^i, z^i) - h_b^i\right|^2.$$

Thus $MSE_c, MSE_0$ and $MSE_b$ penalize the neural network on the residuals of the predicted solution of the PIDE, the initial conditions, and boundary condition respectively. $N_0$ and $N_b$ are the number of labeled and boundary coordinates respectively.

While the temporal collocation points in the CFD-PINN should be linearly spaced for the proper operation in the Fourier domain, we want to emphasize that there is no relationship between the number of labeled coordinates and the number of collocation coordinates, an important advantage inherited from the PINN framework. The number of collocation coordinates is purely chosen digitally and can sample the domain to arbitrary accuracy while labeled coordinates must be physically measurable or definable. The choice to increase the temporal collocation only applies computational complexity to the CFD-PINN but does not limit any of the experimental requirements. Thus, the CFD-PINN is intrinsically compatible with compressive sensing and the collected labeled data can be significantly down sampled compared to the collocation without any requirement to interpolate the data to adequately predict the complete spatiotemporal domain.

To emphasize this point, in Fig. 1, the 64 labeled coordinates fed to the CFD-PINN were randomly sampled over a domain of 10 pulse widths of the input (500 fs). The selected labeled data is illustrated in Fig. S3. The networks performs equally well as when the labeled data is linearly sampled (Supplemental Fig. S2). We digitally selected 512 linearly spaced temporal collocation coordinates to sample our domain at under 1 fs to finely sample the Raman response function which, in silica, can have a frequency response over 40 THz. We trained the network for over 20,000 epochs with the loss function evolution plotted in Fig. S3b.

Fig. 1.b, illustrates the final prediction of the pulse propagation in time and frequency at 204,800 coordinates (1024 temporal and 200 spatial). We compared the prediction accuracy of the temporal and frequency domain, illustrated in Fig. 1.c, to a machine precise solution found using a standard split step Fourier method (SSFM). The complete spatiotemporal prediction has an $L_2$ norm error, as defined in the methods, compared to this "ground truth" of only 2E-4. The ground truth is plotted over the temporal prediction in Fig. 1.b. The extended performance metrics, including the absolute error, network parameters, labeled data, and training loss, are plotted in Fig. S3. Even with an extremely sparse representation of the input pulse (~6 labeled data point per pulse width), the networks convergence onto the hidden solution in time and frequency is extremely robust. Fig. 1.d. shows that the RMS error over the propagation, as defined in the methods, of the prediction degrades only slightly to 0.0005 as the second order soliton reaches its highest peak power where the pulse is dominated by the nonlinear response. The prediction inaccuracy as a function of the propagation distance behaves similarly to the evolution of the peak power, illustrated in Fig 1.d, and importantly does not accumulate linearly as one would expect in sequential modeling.

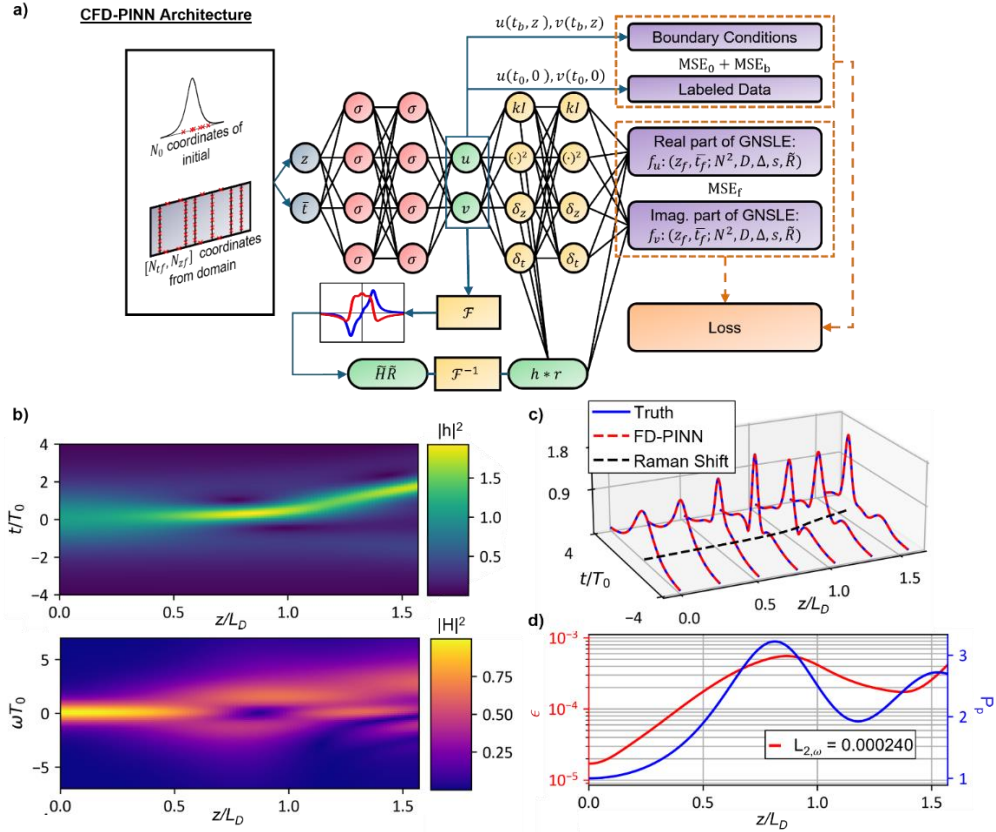

**Fig. 1: Continuous Fourier Domain PINN-** a) The architecture of a CFD-PINN, the collocation coordinates, labeled coordinates, and boundary coordinates, are fed to the network during training to train the network on the physics, labeled data, and boundary data respectively. The trained network can predict b) the temporal (top) and frequency (bottom) field over the full spatiotemporal domain. Compared to c) the ground truth. d) The RMSE error is compared to the peak power of the propagation illustrating non-sequential error accumulation

### 3.  Predictive Denoising in Fourier Domain Physics Informed Neural Networks

A key advantage of the CFD-PINN architecture is its global optimization of the hidden solution. The gradient descent nature of the training process fits the unique solution to the provided physics allowing for arbitrary accuracy at any point in the domain. Even with low fidelity labeled data, this global optimization and physical constraints, provided by the governing equation, enables the model to handle unphysical data points, such as infinite gradients caused by noise. Consequently, the algorithm can decouple input data from noisy perturbations, recover labeled data from the input, and make highly accurate predictions over the spatiotemporal domain.

In Fig. 2, we trained the network on 64 labeled data points which we corrupted with uncorrelated Gaussian amplitude noise with an energy equivalent to 25% of the soliton energy, corresponding to an SNR of 4. We trained the network for 20,000 epochs and then used the trained model to predict over the complete domain as was done in Fig. 1. In Fig. 2.a, the blue points represent the noisy temporal and frequency labeled data fed to the network, while the red and orange curves show the real and imaginary predictions once the model has been trained. The denoising feature of the CFD-PINN architecture is most noticeable here, where a direct comparison can be made between the labeled data, predicted solution, and ground truth. The prediction of the pulse propagation at 204,800 coordinates is illustrated in Fig. 2.b. Compared to the ground truth, the $L_2$ error is only 0.02 over the full domain and remains almost indistinguishable from the noiseless case. Remarkably, the predicted model not only agrees well with the ground truth at the output, indicating its ability to parse the noisy labeled data, but also increases the input SNR by nearly two orders of magnitude. We trained a variety of CFD-PINNs to predict the spatiotemporal profile with varying amounts of additive noise. The performance of these models, as a function of the labeled data SNR, is illustrated in Fig. 2.c. In Fig. 2.d, the RMS error is plotted against the propagation distance which, similarly to the noiseless case, does not accumulate as a function of z. These features are mainly contributed to the physics during training allowing for the network to find the solution amongst the noise.

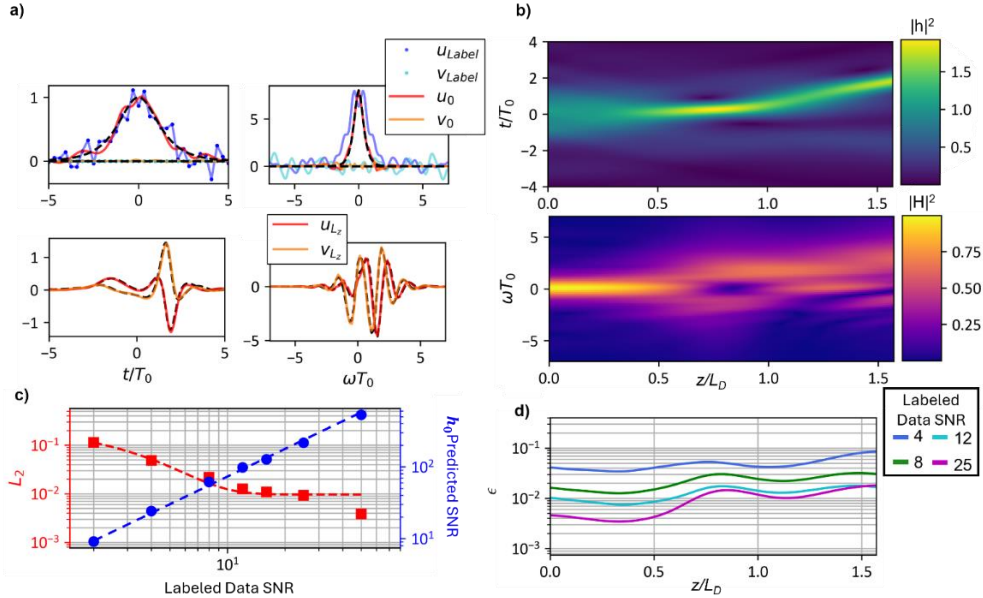

**Fig. 2: CFD-PINN Data Recovery from Noise** – a) The real (blue dots) and imaginary (cyan dots) labeled data corrupted with amplitude gaussian noise equivalent to 25% of the pulse energy. The prediction at the input and output are plotted for the real (red) and imaginary (orange) overlayed upon the ground truth signal (black-dashed). b) the full field prediction in time (top) and frequency (bottom) for the network trained on the labeled data in a. c) the $L_2$ error (top) of the prediction is plotted against the labeled data SNR and the recovered SNR at the input is plotted against the labeled data SNR (bottom). d) the SNR as a function of propagation is calculated as a function of propagation distance.

## 4. Discovery of Hidden Continuous Physics from Discrete Fourier Domain Physics Informed Neural Network

Previous, PINN parameter estimation in ultrafast optics[38–40] relied on continuous time modeling which necessitates the need for scattered data sampled across a full spatiotemporal domain. However, generally speaking, ultrafast optical field data can only be captured at discrete points in propagation. To remedy this, we introduce the discrete FD-PINN (DFD-PINN), similarly to the CFD-PINN, incorporates both temporal and Fourier domain physics into the loss function to efficiently train the network in the data-starved and noisy regime.

Unlike the CFD-PINN, labeled coordinates are physically connected through an unknown governing equation. We achieve this by discretizing the spatial domain and hidden solution through a Runge Kutta method (RKM). This numerical method allows the network to link an input and output pair through discrete steps of propagation. Through the training process, the network discovers the physical model by minimizing the error between the predicted input and output fields and the labeled data. The DFD-PINN is particularly targeted at discovering physics from spatially separated data points, i.e. the labeled coordinates exist only at discrete locations, input and output.

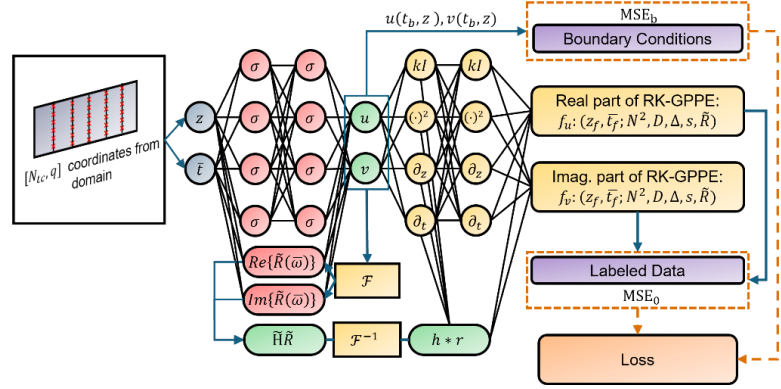

**Fig. 3: Discrete Fourier Domain PINN-** The complete training architecture for the DFD-PINN for nonlinear response discovery. The spatial domain is sliced into q-collocated spatial coordinates each comprising of $N_{tc}$ temporal coordinates. The hidden layers of the RKM are predicted at these coordinates. The training process discovers the Raman response by minimizing the residual on the governing equation.

We built a DFD-PINN by representing the hidden layers of a q-order implicit RKM by a neural network. The governing equation can be expressed in a single RKM step of order q by the real, $u$, and imaginary, $v$, components at the input, $z = 0$ and at the output point, $z = \delta z$ by the set of equations:

$$\begin{aligned}
u_q(\bar{t}, 0) &= u(\bar{t}, z_q) + \delta z \left( F'_u(\bar{t}, z_q)(\alpha_q) \right) \\
v_q(\bar{t}, 0) &= v(\bar{t}, z_q) + \delta z \left( F'_v(\bar{t}, z_q)(\alpha_q) \right) \\
u_q(\bar{t}, \delta z) &= u(\bar{t}, z_q) + \delta z \left( F'_u(\bar{t}, z_q)(\beta_q - \alpha_q) \right) \\
v_q(\bar{t}, \delta z) &= v(\bar{t}, z_q) + \delta z \left( F'_v(\bar{t}, z_q)(\beta_q - \alpha_q) \right).
\end{aligned} \qquad (5)$$

Here, $F'_u = F_u - u_z$ and $F'_v = F_v - v_z$. $h_q = u_q + iv_q$ is the corresponding qth prediction at either the input or output. $\beta$ and $\alpha$ are the butcher table coefficients, and $z_q$ are the spatial locations of the hidden layers of the RKM.

The DFD-PINN neural network architecture is shown in Fig. 3. The collocation coordinates are now specific coordinates representing the spatial location of each hidden layer in each stage. Each hidden layer contains $N_{tc}$ temporal coordinates. The neural network has an output size

$N_{tc}$ by $q$ which predicts each hidden layer of the RKM. To allow the network to learn physical parameters, we define an array of trainable variables in the temporal domain to represent the Raman response of the governing equation.

During the training process, physics is enforced by taking each predicted hidden layer and calculating the propagated (back-propagated) field using the governing equation and comparing it to the output (input) labeled data. The network is optimized by minimizing the loss function:

$$Loss = MSE_0 + MSE_{\delta z} \tag{6}$$

$$= \frac{1}{N_{tc}q} \sum_{j=1}^{N_0} \sum_{i=1}^{q} |h_q(t^j, z_i) - h_0^j|^2$$

$$+ \frac{1}{N_{tc}q} \sum_{j=1}^{N} \sum_{i=1}^{q} |h_q(t^j, z_i) - h_{\delta z}^j|^2.$$

Through this training process the network weights, including the trainable variable defining the Raman response, are updated simultaneously. Introduction of the RKM solver to the algorithm necessarily will lead to intrinsic discretization error. However, this can be mitigated by selecting a RK order that introduces a discretization error well below the machine precision as described in section 7 of the supplemental document.

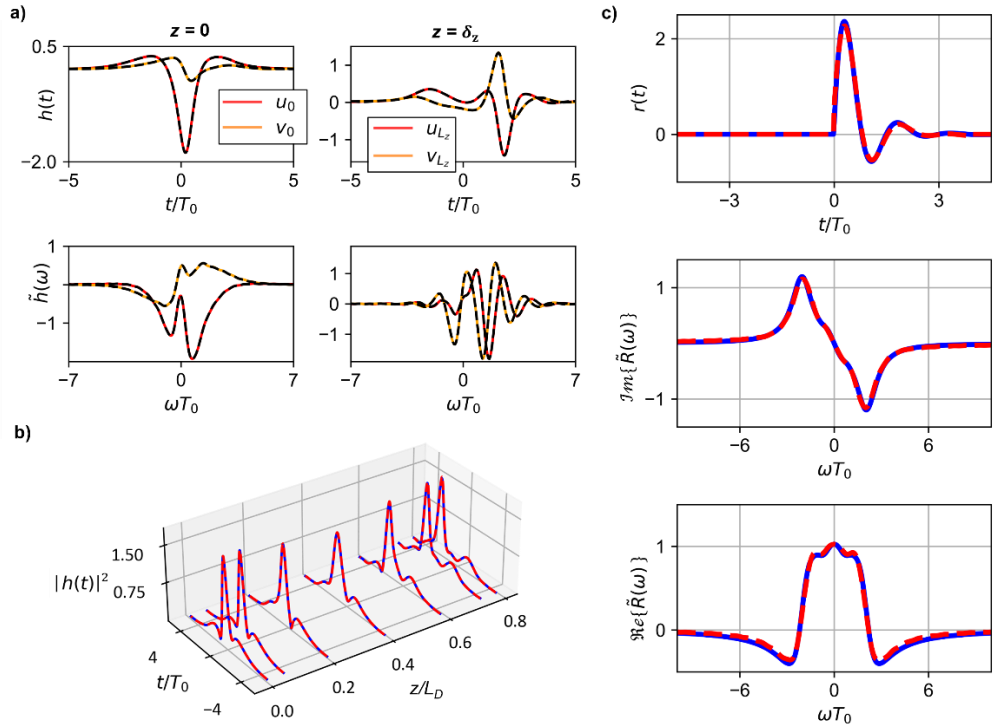

**Fig. 4: Raman response discovery-** a) Labeled data points at the input and $z = \pi/4$ for the real (blue) and imaginary (cyan) compared to the discovered real (red) and imaginary (orange) input and output. b) the discovered (red-dashed) propagation between the input and output via the hidden RK compared to the truth (blue). c) the discovered Raman response (red-dashed) compared to the truth (blue) in the time domain and frequency domain

In Fig. 4, the DFD-PINN was trained to discover the Raman response from labeled data consisting of an input and output pair over half of a soliton period. We obtained this data by simulating the soliton propagation from its shortest pulse width propagated over a distance of $\pi/4$, using a SSFM. The choice to use half the propagation was made to shrink the required number of RK stages and improve stability while ensuring that the introduced discretization error is still well below machine precision. However, the Raman response can still be found with a high fidelity if a longer propagation is considered, more hidden layers are included, and a longer training is conducted (see Fig. S4).

The DFD-PINN was implemented using a 4-layer DFD-PINN consisting of 100 neurons each. We used a RK-order of 200 stages which introduces a discretization error around 1E-42, well below our machine precision. The hidden layers of the 200 stage RKM were predicted at 512 linearly spaced temporal coordinates over a time bandwidth of 12 pulse widths. This corresponds to a physical sampling of around 1.17 fs, an experimentally achievable sampling for standard FROG systems. A soft boundary was also included in the time domain to ensure stability during the training process when the Fourier transformation is taken.

The network was trained over 5,000 epochs. Fig. 4.a illustrates the average of the input and output that the network recovered from the discovered RK hidden layers. The agreement between the truth, in black, and predicted data, in red and orange, indicates a robust discovery of the hidden layers and a recovery of the propagation between the input and output pair. This is also illustrated by the great agreement between the RK layers and truth illustrated in Fig. 4b, a complete spatiotemporal representation of the discovered propagation and further details of the DFD-PINN parameters are provided in Fig. S5.

The agreement between the true propagation indicates the network discovered the Raman response from these two spatial slices with a high level of precision. The discovered response is compared to the truth in Fig. 4.c in red and blue respectively in the time domain as well as Fourier domain. The discovered response has 0.0047 RMSE in the time domain. Importantly, even temporal locations in the response that only have a minor effect on the pulse dynamics can be discovered with a high fidelity such as the minor lobes in the Raman response at $t/T_0 > 1$.

Unlike the CFD-PINN, the collocation coordinates are directly linked to labeled data coordinates via the RKM hidden layers. Therefore, the number of temporal collocation coordinates is directly related to the number of labeled data points required for training. This also implies that the choice of sampling in your labeled data is directly linked to the sampling of the physics of your system.

## 5. Hidden Continuous Physics Discovery from Noisy Data

Similarly to the CFD-PINN, a key advantage of the added physics in the architecture allows for global optimization of the hidden RKM stages. In addition, the added RKM discretization and physical link between the input and output pairs allows the network to parse uncorrelated points between the input and output labeled data, such as gaussian noise, which are physically impossible by the provided PIDE. Consequently, the algorithm can decouple input data from noisy perturbations, recover labeled data from the input, and make highly accurate reconstructions of governing physics without prior knowledge.

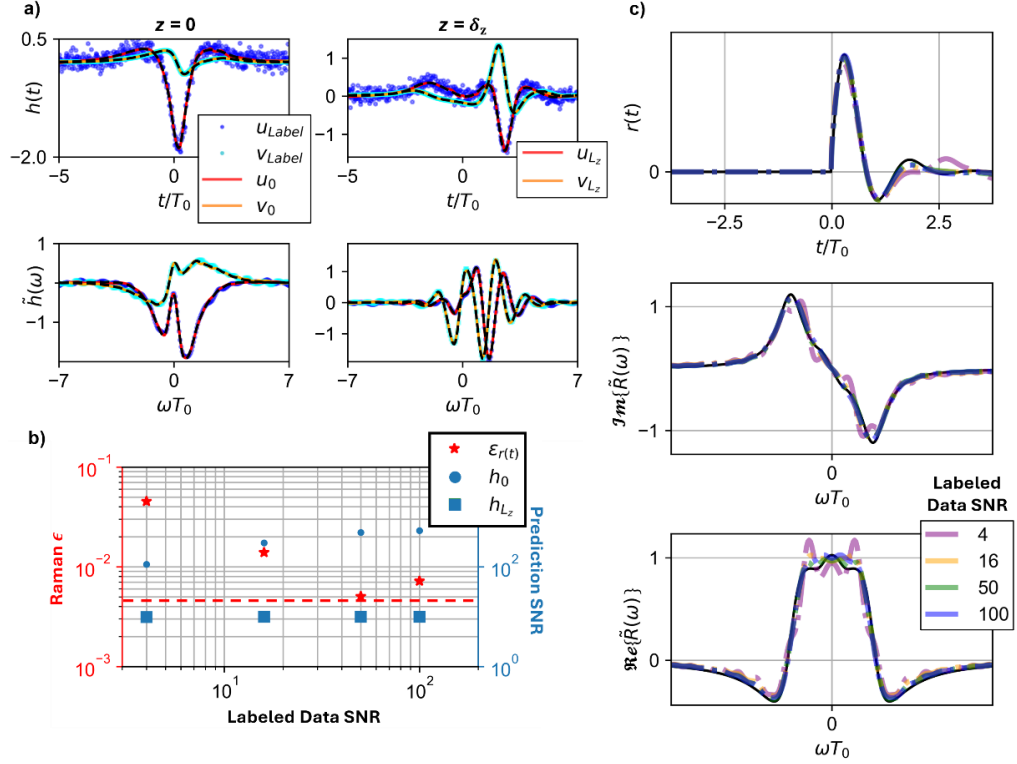

**Fig. 5:** a) Labeled data points at the input and $z = \pi/4$ for the real (blue) and imaginary (cyan) compared to the discovered real (red) and imaginary (orange) input and output for an input SNR of 8. b) the Raman response of a variety of labeled data SNR compared to the truth (black) for the time (top) and frequency domains (bottom) c) the RMSE error (red-star) of the discovered Raman compared to the noiseless case (red-dashed line) and the recovered input (blue circle) and output (blue square) SNR of the labeled data.

In Fig. 5, we added noise to the labeled data in varying degrees of severity and trained multiple networks to recover the Raman response. Fig. 5a illustrates the recovered input and output of the field, in red and orange, compared to labeled data, in blue and cyan, corresponding to an SNR of 16. The truth is illustrated as the black dashed line. As illustrated in Fig. 5b, by the red stars, the error of the RMSE remains below 0.05 and recovered SNR of the labeled data increases as the labeled data SNR degrades. The Raman response is discovered with a high fidelity, below 0.05 error for all cases, and approaches the error of the noiseless case, illustrated as the red dashed line, when the SNR is greater than 20.

The recovered Response functions are illustrated in Fig. 5.c, for a variety of labeled data SNR. Stronger oscillations in the real and imaginary components of the frequency response become apparent when the noise gets to be around 25% of the total soliton energy (purple line). This can be explained by the network overfitting the Raman function to the noise of the labeled data. The minor lobe of the Raman response becomes buried under the noise and the propagation physics begins to overfit to the noise in the labeled data. In theory this drawback could be mitigated by including further constraints on the response function such as fitting the parameters of an equation or introducing further constraints in the loss function.

## 6. Discussion and Conclusion

By treating collocation points as a temporal series, our FD-PINN leverages both physical and Fourier domain information to efficiently train neural networks to predict solutions and discover parameters of a broad class of equations. The inclusion of governing physics allows the FD-PINN to parse noisy data and recover true data from unphysical values from low SNR labeled data.

The CFD-PINN is designed to be remarkably robust to noisy training data and can effectively parse captured data, improving the labeled data's SNR. The choice of labeled data does not reflect the performance of the network to learn the physics thus accurate prediction can be found even when the labeled data is sparse and random. Moreover, this deep learning method inherently includes the governing physics, ensuring no information is lost. This provides transparency compared to black-box deep learning methods and is most effective at prediction over a large domain with extreme precision and accuracy.

We demonstrated the CFD-PINN architecture by predicting the spatiotemporal domain of a second-order soliton governed by the generalized pulse propagation equation in the data-starved and noisy regime. Our method predicts the spatiotemporal field with over an order of magnitude higher fidelity compared to the provided initial condition. Additionally, the solution of the PIDE is accurately predicted over the complete domain without interpolation of the labeled data. The improvement in SNR provided by the CFD-PINN ensures this algorithm is experimentally robust in the presence of unavoidable noise.

The CFD-PINN offers a highly attractive pathway to mitigate numerical errors and supervision in the modeling process. However, many of the other issues which plague the vanilla PINN architecture have not been addressed by this architecture. Model generalization and the accurate prediction of high frequency components in the optical field, caused by high nonlinearity, are not addressed in our described algorithm. The CFD-PINN can be extended to incorporate the methods proposed by Jiang et, al which can improve the prediction of high frequency components[63] and the generalization of the prediction.[37]

We extended the FD-PINN architecture to demonstrate discovery of continuous physics through a harmonious integration of an implicit RKM with the FD-PINN through the DFD-PINN. We used the DFD-PINN to recover the Raman response from spatially sparse coordinates captured at only two spatially separated propagation locations. With this sparse data, tens to hundreds of hidden parameters can be discovered simultaneously without external supervision. Additionally, the DFD-PINN discovers the unique propagation solution linking the two labeled coordinates and displays excellent robustness to noise. However, this inversion algorithm is slightly more susceptible to overfitting. Future work should extend this method to include additional physical constraints to mitigate overfitting issues by the inclusion of multiple RK-steps, reducing the number of fitting parameters, or further constraints in the loss function.

Our new intelligent discovery and prediction algorithm will be highly impactful for high-dimensional physics, where tens to hundreds of parameters must be fit, or for predicting complex low SNR dynamics. We envision our methods will be best suited to tackle forward and inverse problems with sparse available data or noisy initial conditions, common in experimental conditions. Our easily implementable and simple machine learning architecture ensures high fidelity predictive modeling and hidden physics recovery for experimental applications such as image reconstruction, pulse characterization and shaping, as well as hidden parameter discovery.

**Funding.** Office of Naval Research (N00014-19-1-2251).

**Acknowledgments.** J.M. acknowledges the NDSEG fellowship from the Air Force Research Laboratory.

**Disclosures.** The authors declare no conflicts of interest.

**Data availability.** Data underlying the results presented in this paper are not publicly available at this time but may be obtained from the authors upon reasonable request.

# FOURIER DOMAIN PHYSICS INFORMED NEURAL NETWORK: SUPPLEMENTAL DOCUMENT

## 1. Generalized Pulse Propagation Equation Formulation

In expressing Eq 2 we have made the following normalizations to express the system in dimensionless parameters:

$$h = \frac{U}{\sqrt{P_0}}, z = \frac{Z}{L_D}, t = \frac{T - v_g}{T_0}, N^2 = \frac{L_D}{L_{NL}}, L_D = \frac{T_0^2}{|\beta_2|}, L_{NL} = \frac{1}{\gamma P_0}, s = \frac{1}{\omega_0 T_0}, \Delta = \frac{\beta_3}{6|\beta_2|T_0}.$$

Where $P_0$ is the input peak power, $T_0$ is the pulse width, $\omega_0$ is the pulse frequency in radians, $\gamma$ is the nonlinear parameter, $\beta_2$ is the material dispersion, $v_g$ is the group velocity at $\omega_0$, and $L_D$ and $L_{NL}$ are the dispersive and nonlinear lengths respectively.

To convert to physical units, the parameters for the trained networks discussed in the manuscript are:

| Parameter | Definition | Value |
| --- | --- | --- |
| $\beta_2$ | Group Velocity Dispersion | 23 fs²/mm |
| $\beta_3$ | Third Order Dispersion | 30 fs³/mm |
| $\gamma$ | Effective nonlinear coefficient | 0.004 1/Wm |
| $T_0$ | Pulse Width of Input Soliton | 50 fs |
| $\omega_0$ | Carrier Frequency | $2\pi \times 359$ THz |

The peak power, $P_0$, is adjusted accordingly to determine the soliton number.

The last terms on the right-hand side of Eq 2 represents the Kerr-nonlinearity and Raman response of the system. For silica fibers, the Raman contribution is well approximated by

$$r(t) = (1 - f_b)T_0 \frac{(\tau_1^2 + \tau_2^2)}{\tau_1 \tau_2^2} e^{\frac{T_0 t}{\tau_2}} \sin\left(\frac{tT_0}{\tau_1}\right) + f_b T_0 \left(\frac{2\tau_b - t}{\tau_b^2} e^{\frac{T_0 t}{\tau_b}}\right)$$

Where the Raman response is parameterized by the fractional molecular contribution, $f_b = 0.21$ and constants $\tau_1 = 12.2\text{fs}, \tau_2 = 32\text{fs}$, and $\tau_b = 96\text{fs}$.

For implementation in the FD-PINN model the GPPE is split into its real and imaginary components by defining $\boldsymbol{h} = \boldsymbol{u} + i\boldsymbol{v}$. The GPPE is thus split into two coupled equations defined:

$$f_u = \tag{1}$$
$$u_z + \frac{D}{2} v_{tt} + N^2 \left[(1 - f_R)\left(v|(u^2 + v^2)|^2\right) + f_R Im\{C\}\right]$$
$$+ sN^2 \left[(1 - f_R)\left(u|(u^2 + v^2)|^2\right)_t + f_R (Re\{C\})_t\right]$$

$$f_v = \tag{2}$$
$$v_z - \frac{D}{2} v_{tt} - N^2 \left[(1 - f_R)\left(u|(u^2 + v^2)|^2\right) + f_R Re\{C\}\right]$$
$$+ sN^2 \left[(1 - f_R)\left(v|(u^2 + v^2)|^2\right)_t + f_R (Im\{C\})_t\right]$$

$$\boldsymbol{C} = \int_{-\infty}^{\infty} ((1 - f_R)\delta(t') + f_R r(t'))(\boldsymbol{u}^2 + \boldsymbol{v}^2) d\boldsymbol{t'} \tag{3}$$

The variables are the same as defined in section 1 while $\boldsymbol{C}$ now represents the convolution of the solution with the delayed response.

## 2. Quantifying Prediction Error

To quantify the error, we calculated the solution to the Generalized Pulse Propagation Equation using a SSFM with a step size which produces error well below the precision of the machine. This solution we assume to be the ground truth and is denoted $h_T$. The absolute error is calculated by the modulus of the prediction or simulation relative to this ground truth expressed:

$$Er.Abs. = |h_T - h|$$

Where $h$ is taken to be the prediction made by the FD-PINN. Note that there is no normalization in this definition and rather gives a quantitative description of where the maximum error exists in the propagation. To better quantify the amount of accumulated error as a function of the propagation direction we used the RMS error defined.

$$RMSE = \sqrt{\frac{\sum_i^{N_t} |h_T^i - h^i|^2}{\sum_i^{N_t} |h_T^i|^2}}$$

$N_t$ is the number of predicted points on the time axis and we sum only over the temporal axis giving us an error function over the z directions. The SNR of the prediction can be equivalently expressed as the inverse of the RMSE:

$$SNR = \frac{1}{RMSE}$$

Finally, to quantify the total error of the prediction over the full domain we used the $L_2$ norm which gives a single number representation of the discrepancy in the total energy of the prediction compared to the ground truth. We define it here as:

$$L_2 = \sqrt{\frac{\sum_j^{N_z} \sum_i^{N_t} |h_T^{i,j} - h^{i,j}|^2}{\sum_j^{N_z} \sum_i^{N_t} |h_T^{i,j}|^2}}$$

Where now the sum is over both the time and spatial dimensions.

We wrote the algorithm for the FD-PINN using TensorFlow 2.0. We used a custom limited-memory Broyden-Fletcher-Foldfarb-Shanno algorithm (LBFGS) and full-batch gradient descent to optimize the weight and biases of the network. We generated the simulation data by solving the Generalized Pulse Propagation equation using a home brewed split step solver in MATLAB and Python.

## 3. First-moment Raman Response approximation for the Nonlinear Schrodinger Equation

Eq 4 of the main text can be written:

$$\boldsymbol{h_z} - i\frac{D}{2}\boldsymbol{h_{tt}} + \frac{\Delta}{6}\boldsymbol{h_{ttt}} - iN^2(1-f_R)\boldsymbol{h}|\boldsymbol{h}|^2 + N^2(1-f_R)s\big(\boldsymbol{h}|\boldsymbol{h}|^2\big)_t - \tau_R \boldsymbol{h}\big(|\boldsymbol{h}|^2\big)_t = \boldsymbol{0}$$

where $\tau_R = T_R/T_0$ and $T_r$ is the first moment of the Raman response function. Fig. S1 illustrates the difference between the GNLSE approximation and the generalize pulse-propagation equation (Truth).

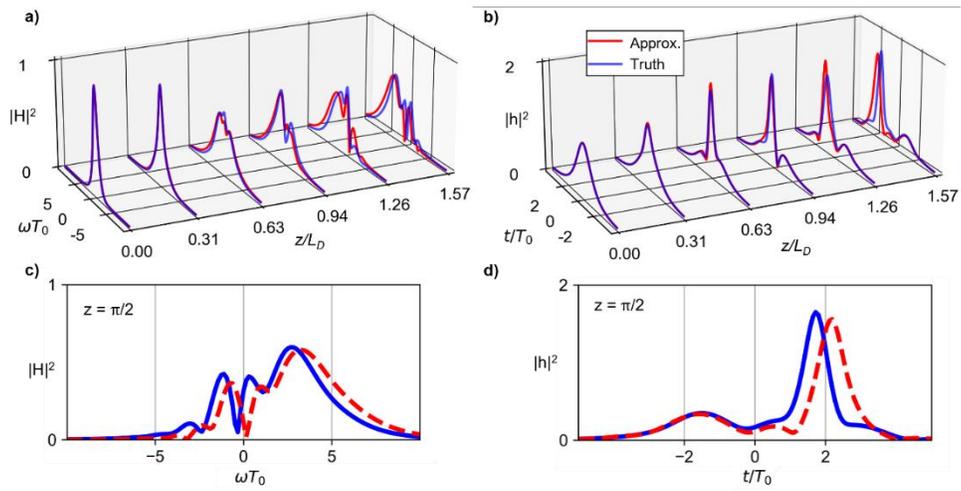

Fig. S 1: **Error of First Moment Approximation** The a) frequency and b) time domain predictions for the first moment approximation (blue) compared to the ground truth prediction (red) both are solved to machine precision using a standard SSFM. The accumulated error of the approximation compared to the truth over one soliton period in c) frequency and d) time.

## 4. CFD-PINN linearly Sampled Labeled Data

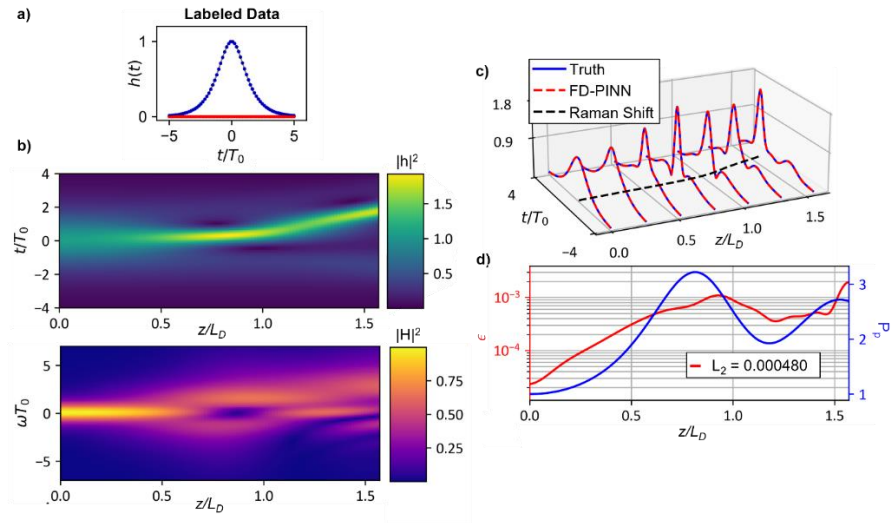

Fig. S 2: **Continuous Fourier Domain PINN Linearly Sampled Input** a) The input points for the real (blue) and imaginary (red) for the labeled data. b) the temporal (top) and frequency (bottom) prediction over the full spatiotemporal domain. Compared to c) the ground truth. d) The RMSE error is compared to the peak power of the propagation illustrating non-sequential error accumulation

## 5. CFD-PINN Training and extended performance metrics

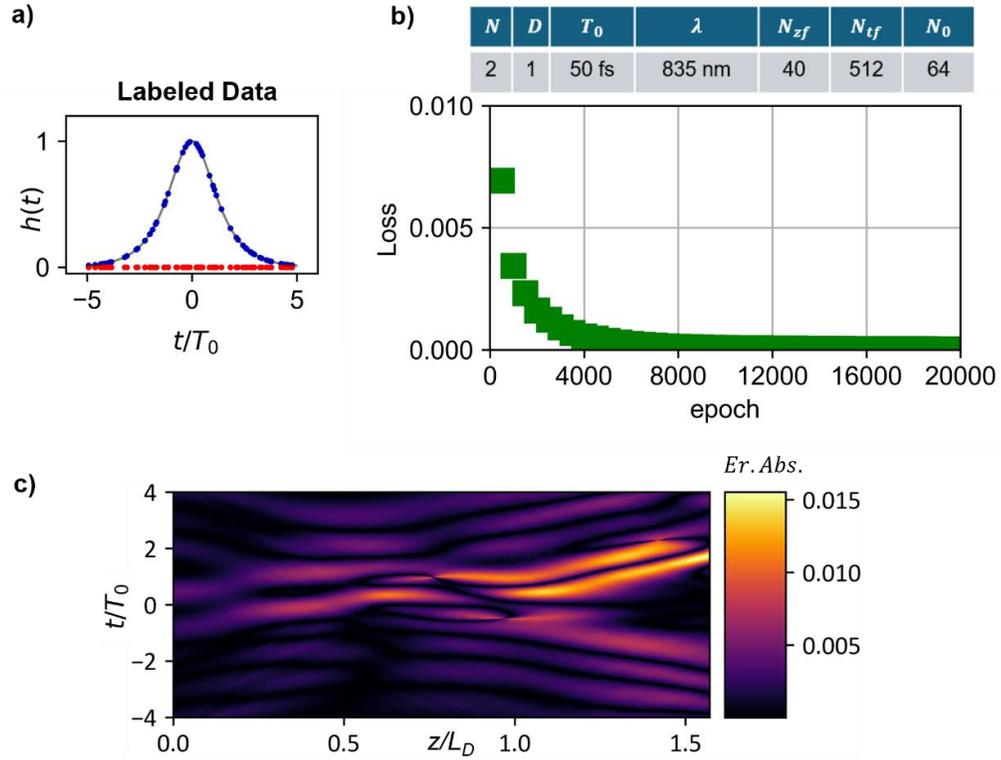

Fig. S 3: **Extended CFD-PINN performance metrics and parameters** a) The input points for the real (blue) and imaginary (red) for the labeled data used for Fig.1 of the main text. b) table of parameters (top) used for the GNLSE to compute the residual on the PIDE and the corresponding loss function as a function of epoch (bottom). c) The absolute error the prediction domain

## 6. DFD-PINN Training and Extended Performance

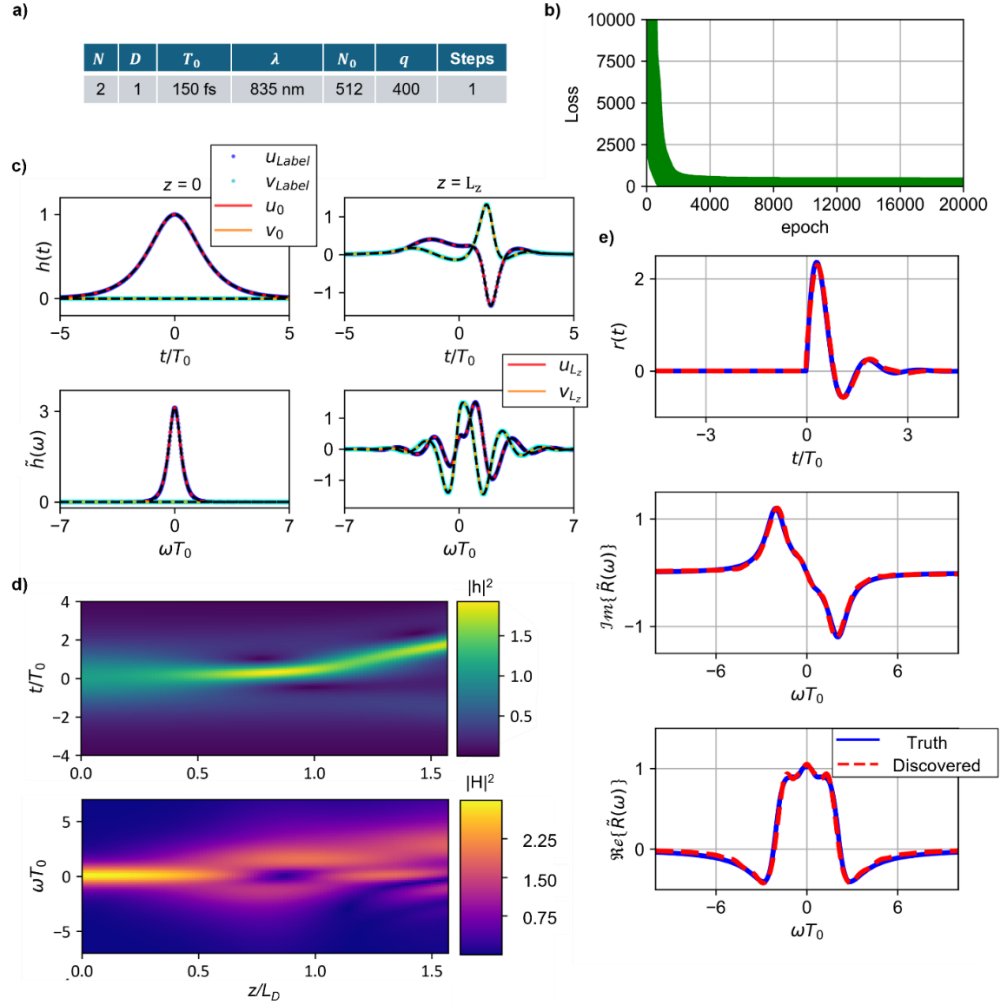

Fig. S 4: **Raman response discovery from $\delta z = \pi/2$** a) Parameters used to train the DFD-PINN and b) the loss as a function of epoch. After training, the c) labeled data points at the input and $z = \pi/2$ for the real (blue) and imaginary (cyan) compared to the discovered real (red) and imaginary (orange) input and output. d) The full propagation discovered by the RK-method and e) the discovered Raman response (red-dashed) compared to the truth (blue) in the time domain and frequency domain

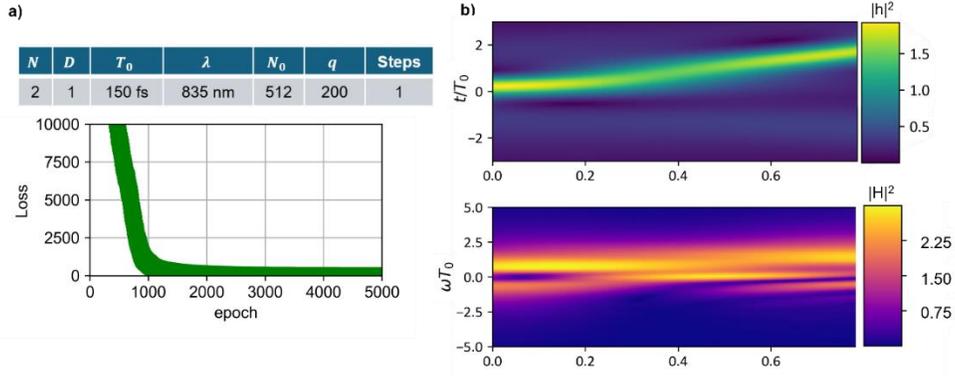

Fig. S5: **Extended CFD-PINN performance metrics and parameters** a) The parameters used to define the network (top) and the loss as a function of epoch (bottom) for the network used to discover the Raman response in Fig. 4 of the main text and b) the 200 hidden stages of the RKM illustrated as a heat map over a propagation distance.

## 7. Mitigating Discretization Error

The RKM order is appropriately selected such that the step size, $\delta z$, does not incorporate numerical error into our training process. To do so the introduced error should exist well below the machine precision. Empirically we can estimate the bound of the truncation error and chose the appropriate RK method using the equation[1]

$$q = \frac{\log(\epsilon)}{2\log(\delta z)}$$

Where $\epsilon$ is the machine precision. For a double precision machine we find $\epsilon = 10^{-16}$. For our choice of $\delta z = \pi/4$ an appropriate RK order of 77 would put our RK method truncation error bound below the precision of the machine. We selected an RK method of order 200, which introduces an error of 1.08E-42 below even 128-bit precision. Importantly the choice to add more orders to the RK method will increase the PINN network complexity linearly while the truncation error will decrease exponentially meaning that the error of this algorithm may easily exist below the machine precision.

### References

1. Iserles, A. *A First Course in the Numerical Analysis of Differential Equations*. (Cambridge University Press, 2009).